# The effect of national and international multiple affiliations on citation impact

Sichao Tong[1], Ting Yue[1,2], Zhesi Shen[1], Liying Yang[1*]

1. National Science Library, Chinese Academy of Sciences, Beijing, 100190, China

2. Department of Library, Information and Archives Management, School of Economics and Management, University of Chinese Academy of Sciences, Beijing 100049, China

*corresponding author: yangly@mail.las.ac.cn

Sichao Tong's ORCID: 0000-0003-0658-9963

Zhesi Shen's ORCID: 0000-0001-8414-7912

**Abstract**

Researchers affiliated with multiple institutions are increasingly seen in current scientific environment. In this paper we systematically analyze the multi-affiliated authorship and its effect on citation impact, with focus on the scientific output of research collaboration. By considering the nationality of each institutions, we further differentiate the national multi-affiliated authorship and international multi-affiliated authorship and reveal their different patterns across disciplines and countries. We observe a large share of publications with multi-affiliated authorship (45.6%) in research collaboration, with a larger share of publications containing national multi-affiliated authorship in medicine related and biology related disciplines, and a larger share of publications containing international type in Space Science, Physics and Geosciences. To a country-based view, we distinguish between domestic and foreign multi-affiliated authorship to a specific country. Taking G7 and BRICS countries as samples from different S&T level, we find that the domestic national multi-affiliated authorship relate to more on citation impact for most disciplines of G7 countries, while domestic international multi-affiliated authorships are more positively influential for most BRICS countries.

## 1. Introduction

Researchers affiliated with multiple institutions are increasingly seen in current scientific environment, e.g., Huang and Chang (2018) show that 87.3% publications are written by multi-institutional authors in genetics and 50.4% in high-energy physics respectively. Hottenrott, Rose and Lawson (2019) find there is a growing trend of multi-affiliated authors, the share is 8% in 1996, and it goes up to 13% in 2018. With direct links with several institutions, a researcher can be consequently recognized as a bridge between institutions, facilitating idea exchange and research collaboration (ESF, 2013; Hottenrott & Lawson, 2017). ESF (2013) also present multiple affiliations is an effective scheme in research collaboration, comparing with general project duration or longer period position, it's more attractive to frontline researchers based on its flexibility. Furthermore, researchers with multiple affiliations are more often found in highly cited publications, regarding to those tested fields and countries, implying their positive influence on scientific impact (Hottenrott & Lawson, 2017; Huang & Chang, 2018; Sanfilippo, Hewitt, & Mackey, 2018). Therefore, studying multi-affiliated researchers' effect on the scientific output of research collaboration is also important, when exploring their influence in facilitating research collaboration.

Naturally, multiple affiliations can also happen among institutions from one country or several countries, whereas current explorations seldom take it into account, in this study, we will classify multi-affiliated researchers by whether they have multiple affiliations within the same country.

From these above, this study will explore scientific outputs with multi-affiliated authorship, among collaborative output, by considering both national multiple affiliations within a same country, and international multiple affiliations within several countries. The following two research questions will be mainly explored in this study:

- Taking scientific output with multi-affiliated authorship as the background, are there any heterogeneities exist by disciplines, or by countries?
- Regarding to the multi-affiliated authorship's effect on citation impact, among collaborative output, are there any difference between multi-affiliated authorship within the same country or from different countries?

## 2. Data and methods

### 2.1 Data

To investigate the effect of multi-affiliated authorship on citation counts in collaboration, we retrieved collaborative publications co-authored by two or more institutions published between 2013 and 2015 with all author address records. Only publications of the Web of Science document types "Article" and "Review" are included in the data collection. A manual institutional disambiguation was undertaken. For citations, we use a 3-year citation window, i.e., for papers published in 2013, the cumulative citations during 2013-2015 are considered.

The ESI classification system is used here to see the different multiple affiliation patterns across disciplines. Publications are categorized into 19 disciplines ("Economics & Business", "Multidisciplinary" and "Social Sciences, General" are excluded), as listed in Table 1. We also aggregate medicine related disciplines, biology related disciplines and engineering related disciplines, respectively, which result in 10 broader science fields.

**Table 1.** The mapping of 19 disciplines considered.

| ***Field*** | ***Discipline*** | ***Discipline (Abbreviation)*** |
|---|---|---|
| Space Science | Space Science | SPA |
| Medicine related | Neuroscience & Behavior | NEU |
| | Psychiatry/Psychology | PSY |
| | Immunology | IMM |
| | Clinical Medicine | CLI |
| | Pharmacology & Toxicology | PHA |
| Physics | Physics | PHY |
| Biology related | Molecular Biology & Genetics | MOL |
| | Biology & Biochemistry | BIO |
| | Microbiology | MIC |
| | Plant & Animal Science | PLA |
| Environment/Ecology | Environment/Ecology | ENV |
| Geosciences | Geosciences | GEO |
| Chemistry | Chemistry | CHE |
| Agricultural Sciences | Agricultural Sciences | AGR |
| Engineering related | Materials Science | MATE |
| | Computer Science | COM |
| | Engineering | ENG |
| Mathematics | Mathematics | MATH |

Countries from Group of G7 (Canada, France, Germany, Italy, Japan, the United Kingdom, and the USA) and BRICS (Brazil, China, India, Russia, and South Africa) are used to compare the differences between multiple affiliations patterns (Table 2).

**Table 2.** Sample countries.

| | ***Country Name*** | ***Abbreviation*** |
|---|---|---|
| ***G7 Group*** | Canada | CA |
| | Germany | DE |
| | France | FR |
| | United Kingdom | GB |
| | Italy | IT |
| | Japan | JP |
| | USA | US |
| ***BRICS Group*** | Brazil | BR |
| | China | CN |
| | India | IN |
| | Russia | RU |
| | South Africa | ZA |

## 2.2 Classification of authorship and publications

For multi-affiliated authorship, there are three types of author in our dataset:

- **NM**: the national multi-affiliated author, who is affiliated with two or more institutions from one country.
- **IM**: the international multi-affiliated author, who is affiliated with two or more institutions from two or more countries.
- **S**: the single-affiliated author.

Given this, we propose a classification to define multiple affiliations at the level of publication, to analyze their scientific output and citation impact in research collaboration:

- P_M: the publication with multi-affiliated authorship. Based on the types of included multi-affiliated authorship, P_M can be classified further,
    - P_NM: the publication with national multi-affiliated authorship.
    - P_IM: the publication with international multi-affiliated authorship.

(It should be noted especially that there is overlap between P_NM and P_IM, namely some publications may have both NM authorship and IM authorship.)

- P_NoM: the publication without multi-affiliated authorship.

In summary, 59.3% of all publications are institutionally collaborative publications, to be considered as the total scientific output of research collaboration in this study. Based on the classification above, Table 3 presents a general overview of the number of publications in different groups. From Table 3, we can see nearly half (45.6%) of the publications include multi-affiliated authorships. There are 35.4% publications out of total publications having NM authorship, which is larger than the share of publications with IM authorship (14.3%).

**Table 3.** Number and share of publications with multi-affiliated authorship.

| | ***Total*** | ***P_M*** | ***P_NM*** | ***P_IM*** | ***P_NoM*** |
|---|---|---|---|---|---|
| ***Pubs*** | 2,137,885 | 976,036 | 755,850 | 305,479 | 1,161,849 |
| ***Share*** | - | 45.6% | 35.4% | 14.3% | 54.4% |

*(Note:* ***Total*** *is total institutionally collaborative publications.* ***Share*** *is share of publications out of total institutionally collaborative publications.)*

### 2.3 Regression analysis

We perform a regression analysis to reveal the effects of multi-affiliated authors on citation counts. Considering the over-dispersed and high skewed citation count data, we utilize a Negative Binomial Regression Model (Hereinafter referred to as NBRM (Bornmann & Daniel, 2006; Bornmann, Schier, Marx, & Daniel, 2012)) with the citation count (TC) of each paper as the dependent variable. To investigate the effect of NM and IM on citations, we use the following two independent variables:

- NM_mark: equals 1 if the paper has at least one NM author; otherwise 0.
- IM_mark: equals 1 if the paper has at least one IM author; otherwise 0.

We also consider several publication-related factors which may affect citation counts as control variables (Peters & Vanraan, 1994; Glänzel, 2001; Aksnes, 2003; Bornmann & Daniel, 2008; Schmoch & Schubert, 2008; Sooryamoorthy, 2009; Persson, 2010; Vieira & Gomes, 2010; Gazni & Didegah, 2011; Didegah & Thelwall, 2013), including the number of institutions, the number of countries, the number of references and the number of authors, for each publication. Those factors associated with citation counts are defined as control variables:

- N_ins: number of institutions
- N_c: number of countries
- N_refs: number of references
- N_a: number of authors.

The Python statsmodels package (Seabold & Perktold, 2010) is used to estimate the regression coefficients. When conducting regression, we only consider publications with 10 or less authors, to avoid the effects of those extra large groups and the possible high correlation between number of authors and number of institutions (Table S2 in SM for number of records used in regression). Variance Inflation Factor test shows there is absence of multicollinearity among these independent variables. For codes and detailed parameter estimations, please refer to the ***Codes availability*** section in ***Supplementary Materials***. We also consider several regression models, and similar phenomena are obtained. For details please refer to the Regression Models section in ***Supplementary Materials***.

## 3. Results

### 3.1 Discipline-based analysis

This section reports the heterogeneities by discipline, involving share of publications and effect on citation impact of collaborative publications, regarding to the national type and international type multi-affiliated authorship.

### 3.1.1 Statistics of multi-affiliated publications

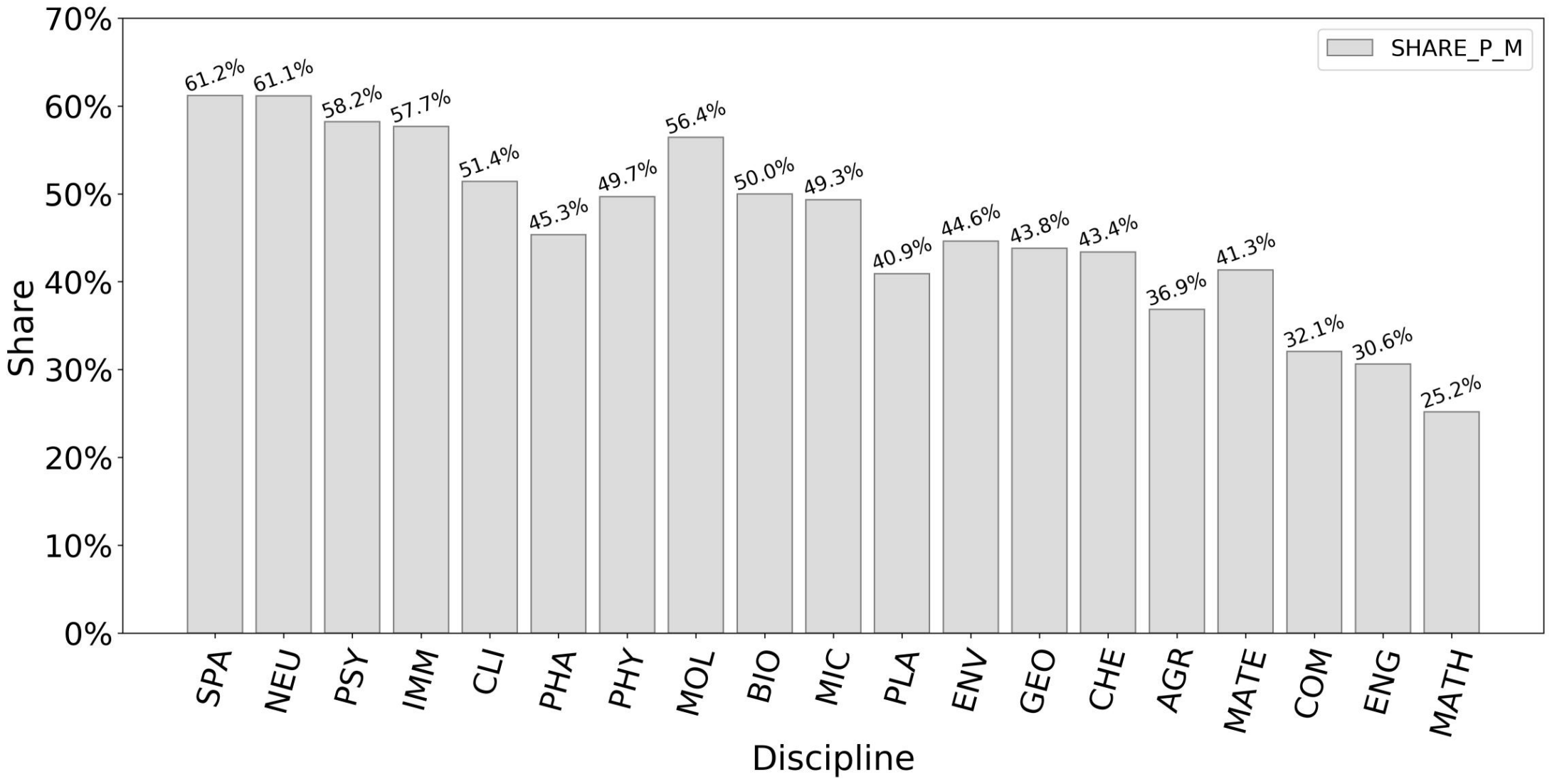

**Figure 1.** Share of publications containing multi-affiliated authorship of each discipline.

Figure 1 shows the share of P_M of given disciplines. From Fig. 1 we can see in research collaboration, SPA, medicine related disciplines and biology related disciplines have relatively higher share of publications with multi-affiliated authorship, while engineering related disciplines and MATH have lower share. 61.2% of SPA's publications contain multi-affiliated authorship, while MATH has the smallest share of multi-affiliated publications (25.2%).

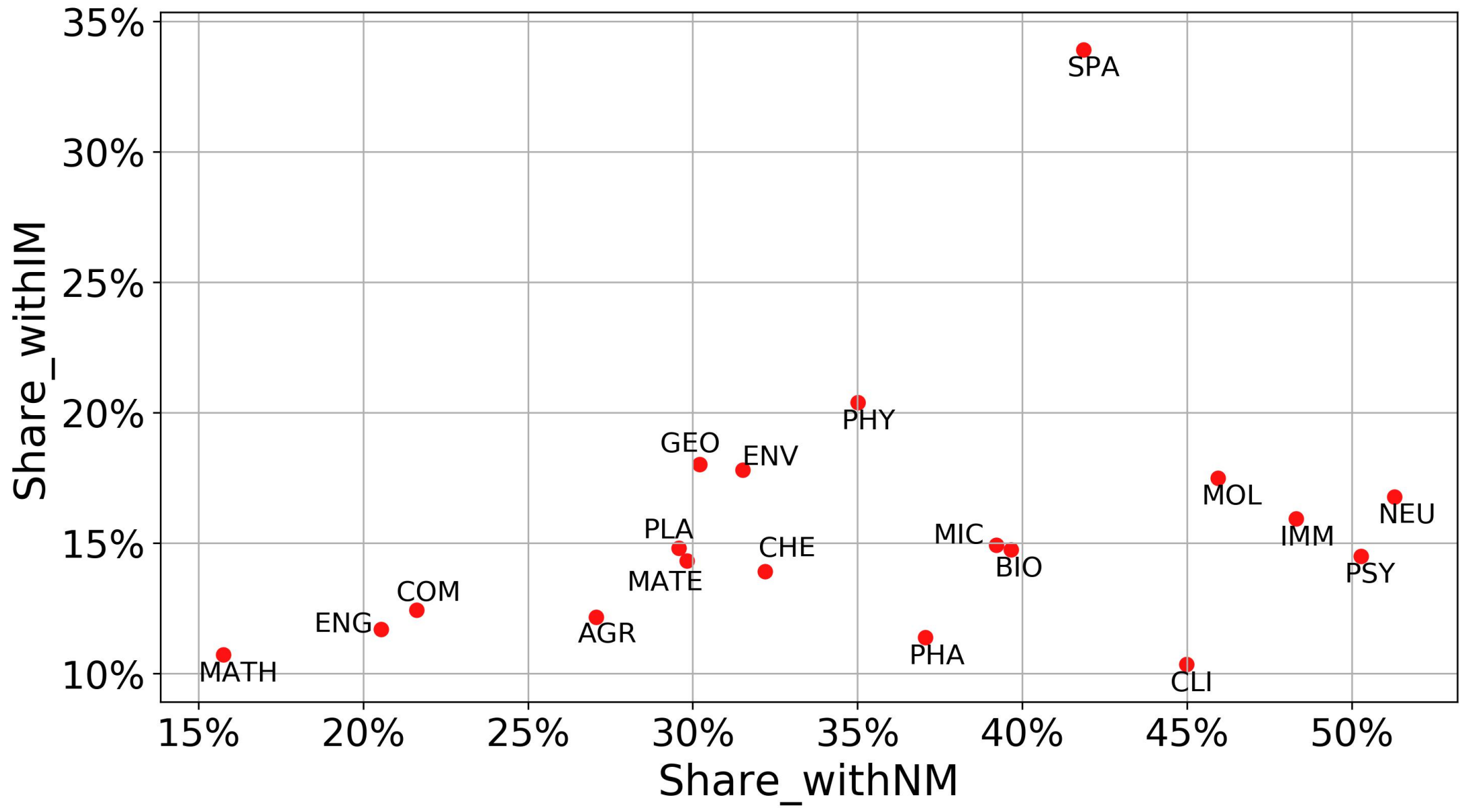

**Figure 2.** Share_P_NM vs. Share_P_IM of each discipline.

Figure 2 visualizes both share of publications containing NM authorship, and share of publications containing IM authorship among research collaboration, in different disciplines. Specific values are provided in ***Table A1*** in ***Appendix***.

We can see that in all disciplines, the shares of P_NM are larger than P_IM. For most medicine and biology related disciplines, there is a big gap (over 20%) between the share of P_NM and P_IM, e.g., in PSY, the proportion of P_NM and P_IM are 50.3% and 14.5%, respectively. For SPA, compared to other disciplines, its share of P_IM is the largest, and the difference between the share of those two types is not significant. Among these disciplines, MATH, ENG and COM have relative smaller shares in both of P_NM and P_IM.

### 3.1.2 Effect of multi-affiliated authorship on citation impact

Previous studies (Hottenrott & Lawson, 2017; Huang & Chang, 2018; Sanfilippo, Hewitt, & Mackey, 2018) indicate that publications which contain multi-affiliated authorship have a larger probability of receiving more citations. Among scientific output of collaboration, we also observe the same phenomenon based on the NBRM result, here we focus on the effect of national and international multiple affiliation on citation impact.

Table 4 presents the results of the NBRM regression showing the size of effects of each observed factor on citation counts. For each independent variable, we report the percent changes in expected citation count for a unit increase in that variable (Long & Freese, 2006). For example, in Chemistry for the variable "NM_mark", the result shows that including national multi-affiliated authorships increases the expected number of citations by about 16.9% when keeping other variables consistent. Please refer to the ***Codes availability*** section in ***Supplementary Materials,*** for the raw regression coefficients.

**Table 4.** Expected percent change of citations affected by each variable, here we mainly analyze national multi-affiliated authorship's effect and international multi-affiliated authorship's effect. We calculate the percent change in expected citation count for a unit increase in each variable.

| ***Field*** | ***Discipline (Abbreviation)*** | **NM_mark** | **IM_mark** | **N_refs** | **N_ins** | **N_c** | **N_a** | **R-Squared** |
|---|---|---|---|---|---|---|---|---|
| ***Space Science*** | ***SPA*** | -2.8 | 4.6** | 1.3*** | 2.8*** | 2.7** | 1.2** | 0.15 |
| ***Medicine related*** | ***NEU*** | 5.9*** | 2.8* | 1.1*** | -0.6 | 13.2*** | 5.2*** | 0.13 |
| | ***PSY*** | 9.3*** | 0.6 | 2.1*** | -0.3 | 12.1*** | 6.8*** | 0.17 |
| | ***IMM*** | 8.0*** | 3.7* | 1.2*** | 1.1* | 11.4*** | 2.4*** | 0.14 |
| | ***CLI*** | 8.9*** | -2.7*** | 1.8*** | 2.7*** | 19.2*** | 7.4*** | 0.14 |
| | ***PHA*** | 9.6*** | 2.1 | 1.1*** | -1.0* | 12.3*** | 4.3*** | 0.14 |
| ***Physics*** | ***PHY*** | 14.7*** | 14.3*** | 2.2*** | -5.8*** | 23.1*** | 6.7*** | 0.13 |
| ***Biology related*** | ***MOL*** | 16.7*** | 13.8*** | 1.3*** | -4.9*** | 14.1*** | 5.8*** | 0.12 |
| | ***BIO*** | 14.5*** | 6.7*** | 1.0*** | -4.5*** | 15.6*** | 3.8*** | 0.09 |
| | ***MIC*** | 5.9*** | 1.2 | 1.4*** | -0.8 | 14.4*** | 3.8*** | 0.16 |
| | ***PLA*** | 9.1*** | 0.9 | 2.2*** | -2.4*** | 17.7*** | 5.4*** | 0.21 |
| ***Environment/Ecology*** | ***ENV*** | 8.8*** | 5.6*** | 1.7*** | -2.0*** | 14.8*** | 6.3*** | 0.14 |
| ***Geosciences*** | ***GEO*** | 3.1*** | 8.2*** | 1.4*** | 0.1 | 9.2*** | 4.9*** | 0.13 |
| ***Chemistry*** | ***CHE*** | 16.9*** | 15.2*** | 1.6*** | -10.9*** | 17.0*** | 8.1*** | 0.16 |
| ***Agricultural Sciences*** | ***AGR*** | 12.4*** | 13.4*** | 2.3*** | -5.9*** | 17.6*** | 6.1*** | 0.15 |
| ***Engineering related*** | ***MATE*** | 15.2*** | 19.3*** | 2.8*** | -12.1*** | 14.7*** | 14.4*** | 0.23 |
| | ***COM*** | 7.1*** | -3.2* | 3.5*** | 0.5 | 30.9*** | 5.1*** | 0.13 |
| | ***ENG*** | 11.1*** | 10.3*** | 3.3*** | -4.2*** | 22.5*** | 5.9*** | 0.16 |
| ***Mathematics*** | ***MATH*** | 4.2** | 13.8*** | 3.9*** | 0.6 | 18.1*** | 4.9*** | 0.11 |

*(Note: *p < 0.05; **p < 0.01; and ***p < 0.001)*

In Table 4, for NM authorship, the values of coefficient in NEU, IMM, CLI, MOL, BIO and COM, are 5.9%, 8%, 8.9%, 16.7%, 14.5% and 7.1%, respectively, while the corresponding values of coefficient for IM authorship are 2.8%,3.7%, -2.7%, 13.8%, 6.7% and -3.2%. We can find that NM authorship's participation is associated with a much greater number of citations, comparing to IM authorship's participation in most medicine related and biology related disciplines, COM as well shows similarly. In PHY and CHE, MATE, ENG and MOL, both NM and IM authorship's participations have significant positive influence on the citations, the value of coefficients are over 10%. In the disciplines of GEO and MATH, IM authorship's participation relates to much more citations, than NM authorship.

### 3.2 Country-based analysis

From the perspective of a certain country, the multi-affiliated authors seem to be more complicated. They can simultaneously have this country's affiliation(s), and other countries' affiliation(s). Accordingly, the multi-affiliated authorship is likely to belong to different types, in terms of those two countries. Therefore, for a certain country, the authorship could be classified into 6 types: NM_Domestic, NM_Foreign, IM_Domestic, IM_Foreign, S_Domestic and S_Foreign. An illustration for the classifications (taking country *A* for example) is shown in Figure 3.

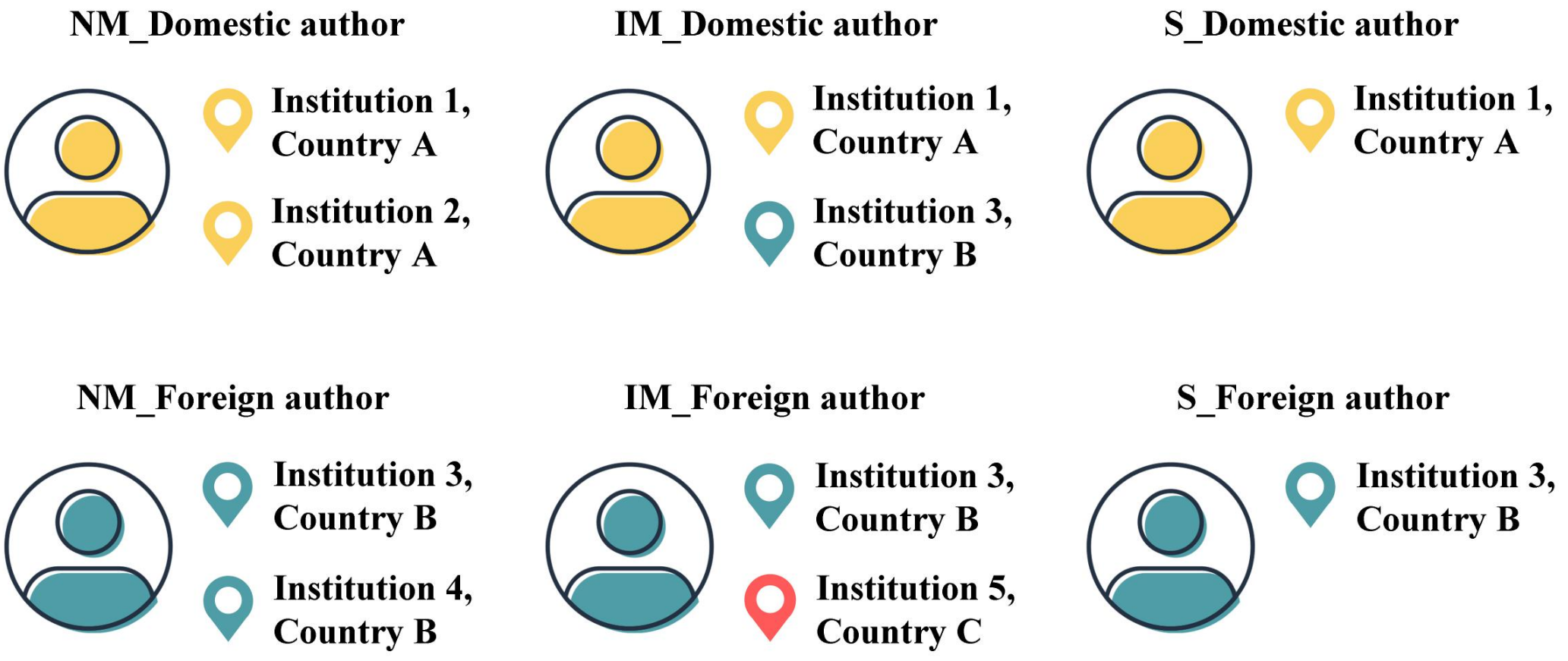

**Figure. 3 Authorship classification for country *A*.**

For different types of authorship (the domestic multi-affiliated authors and foreign multi-affiliated authors), the effect on citation impact may show difference. Here we focus on each country's domestic multi-affiliated authorship. Taking country *A* as an example, it has two kinds of the collaborative publication containing domestic multi-affiliated authorship below:

- P_NM_Domestic: publication with country *A*'s domestic national multi-affiliated authorship.
- P_IM_Domestic: publication with country *A*'s domestic international multi-affiliated authorship.

### 3.2.1 Statistics of multi-affiliated publications

Figure. 4 presents the shares of different collaborative publication groups for the country sample, G7 and BRICS.

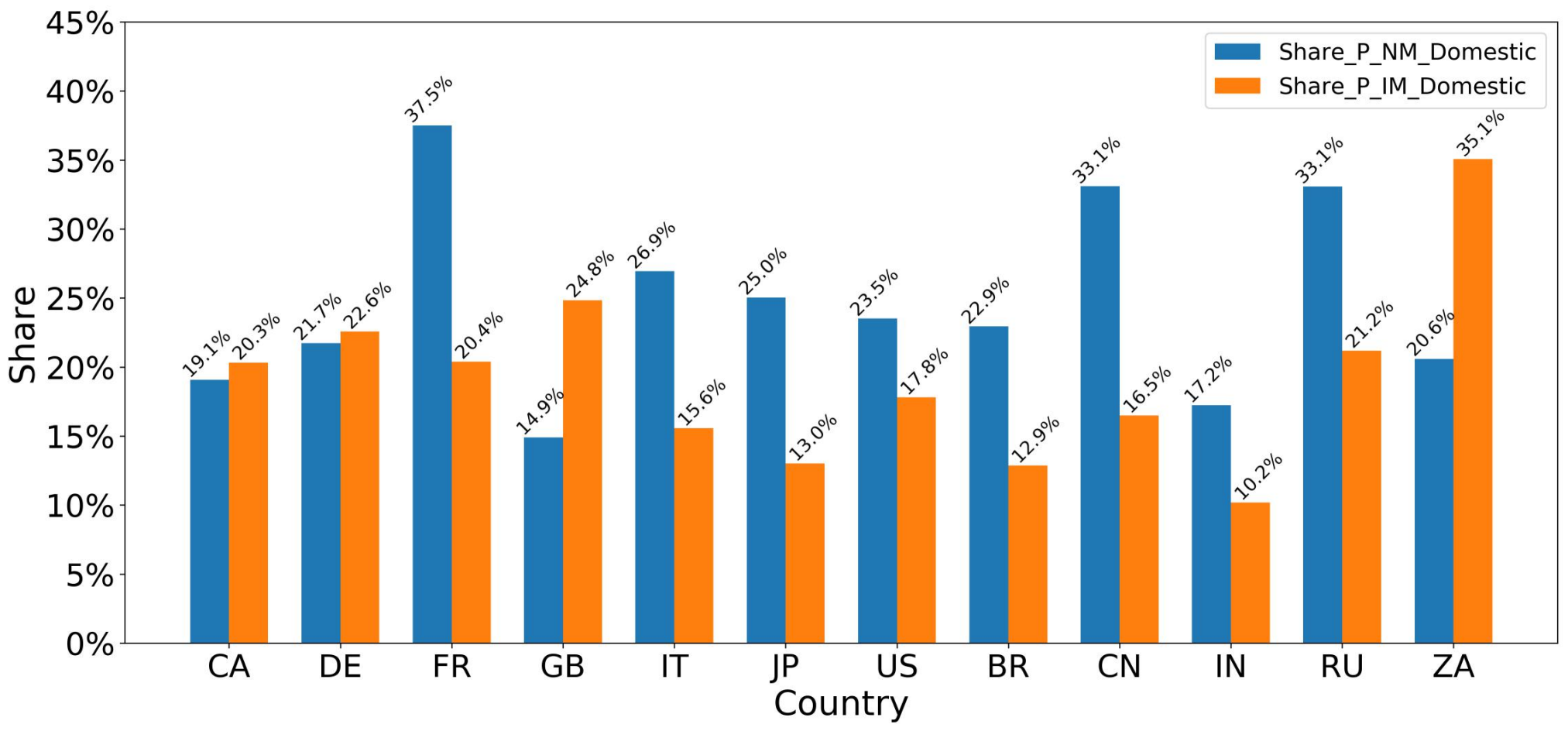

**Figure 4.** Share of different publications groups contain domestic multi-affiliated authorship, for G7 and BRICS countries.

For France, China and Russia, the share of P_NM_Domestic (over 33%) is much higher than other countries, while the value of other countries are basically below 27%. South Africa shows an extremely larger share of P_IM_Domestic, approaching to 35.1%. We investigate the detailed information for the institutions of France. The P_NM_Domestic of France's Top 3 institutions accounts for 42.8%,39.0% and 27.2%, respectively, in the total collaborative publications, which are much higher than other countries (please refer to ***Table A2*** in ***Appendix***).

In the comparison of P_NM_Domestic and P_IM_Domestic for each country, we can see that: the majority of countries have bigger share of P_NM_Domestic, while the United Kingdom and South Africa show extremely larger share of P_IM_Domestic (the United Kingdom: P_IM_Domestic’s share is 24.8% and P_NM_Domestic’s share is 14.9%; South Africa: P_IM_Domestic’s share is 35.1% and P_NM_Domestic’s share is 20.6%). In South Africa and the United Kingdom, the share of P_IM_Domestic for their Top 3 institutions are over 20%, while for other countries, the share for their Top 3 institutions are all below 20% (please refer to ***Table A3*** in ***Appendix***). Canada and Germany show slight difference between the shares of these two publication types.

We then go to the field level to explore the detailed share in different fields, for the G7 and BRICS countries, the two 12 × 19 matrices represented in the figure sets from Figure 5 visualize the normalized share by discipline, of these countries.

**(a)**

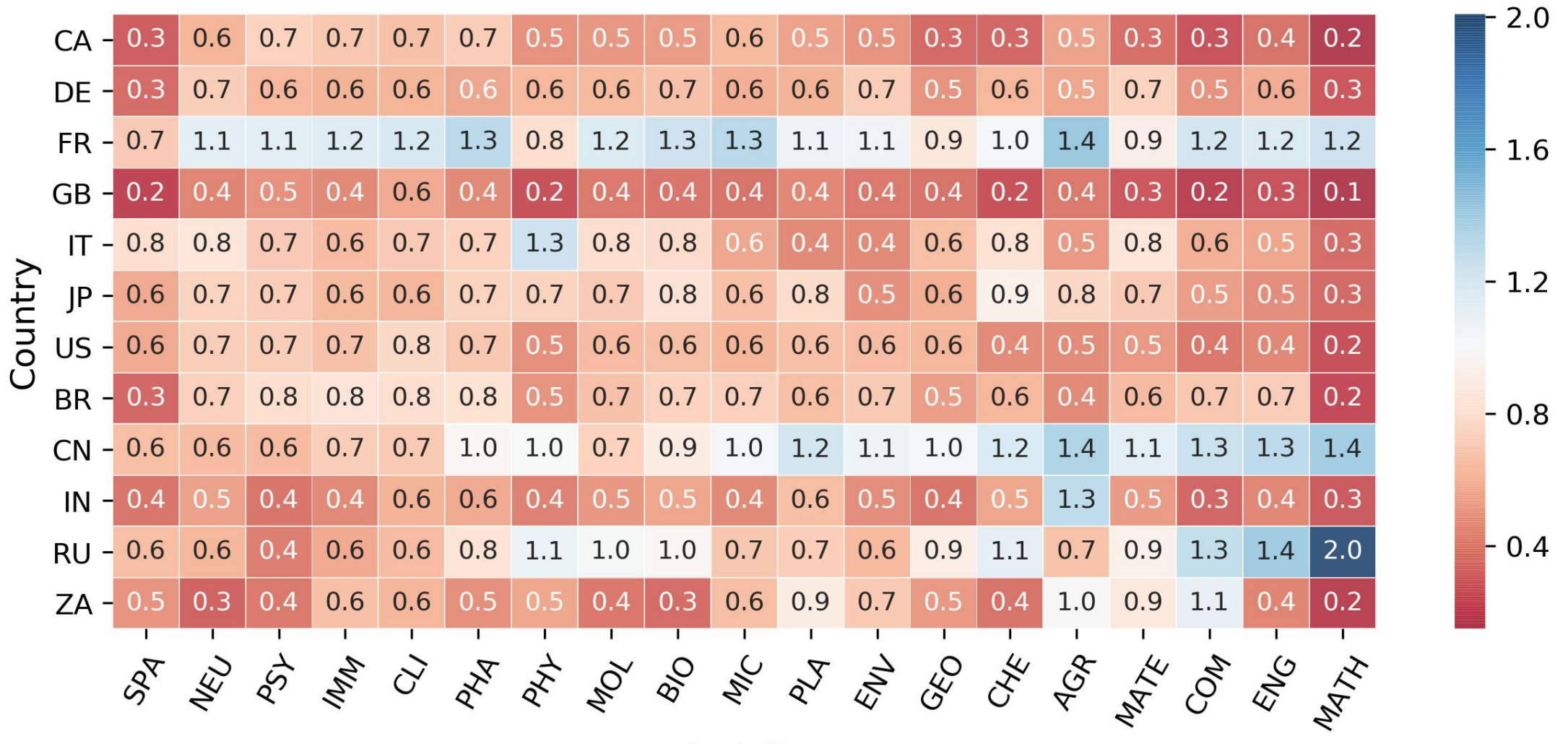

(

b)

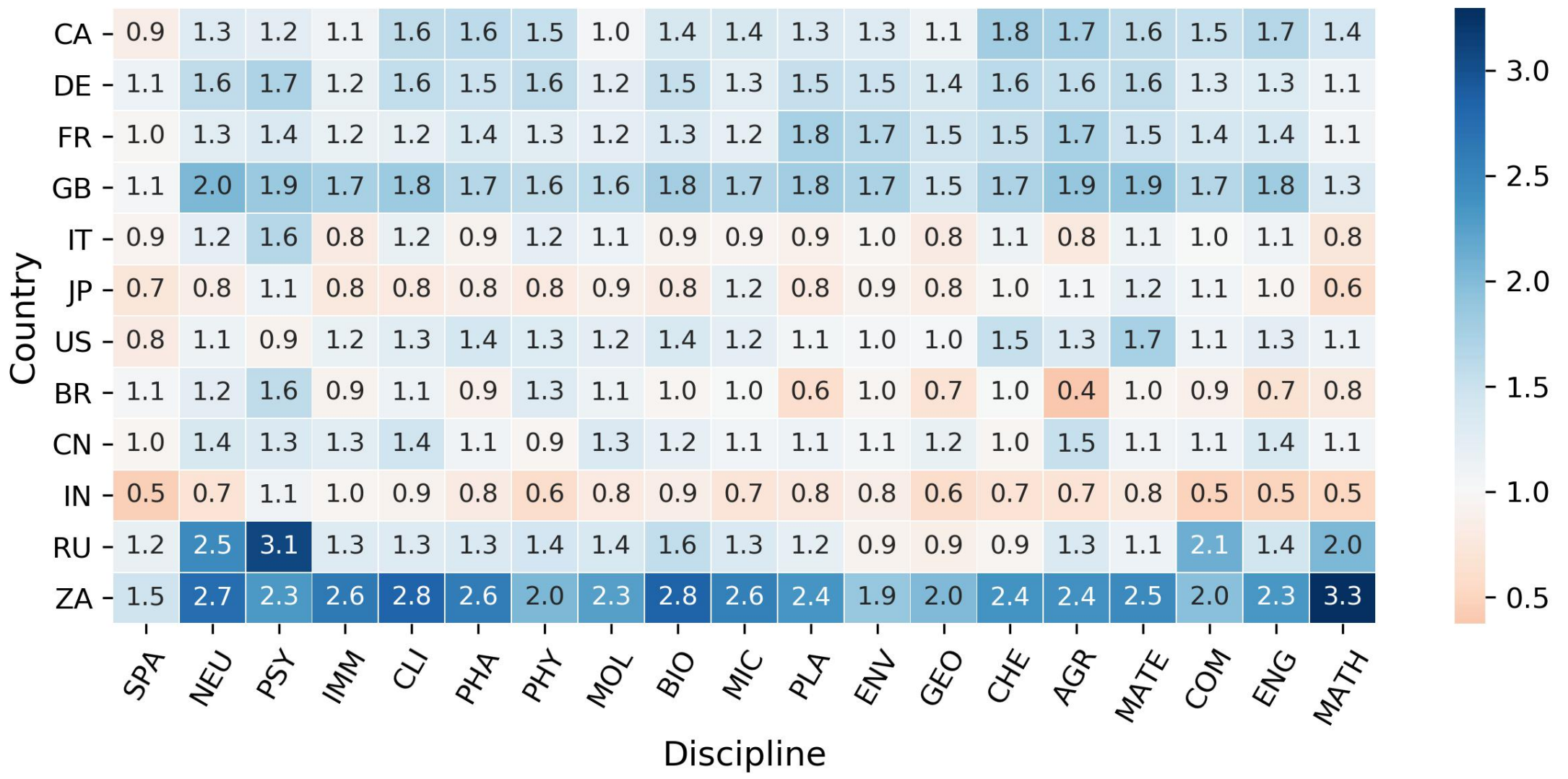

**Figure 5.** Relative share of (a) national multi-affiliated authors and (b) international multi-affiliated authors by discipline, for G7 and BRICS countries. Each square represents the ratio of one country's share of P_IM_Domestic (or P_NM_Domestic) compared with the global baseline of the column discipline. The color of Each square is related to the value of normalized share: the blue color means value is higher than world average, the red color means value is lower than world average, the darker the color is, the larger/smaller the value is. For the original share, please refer to the **Table A4& Table A5** in ***Appendix***.

As shown in Figure 5(a), regarding to P_NM_Domestic, France stands out in the 12 countries, it almost has values larger than 1 among all discipline. China and Russia have more disciplines with normalized share higher than global average, comparing to other countries (except France). They have high value in engineering related disciplines (COM and ENG), PHY and CHE, as well as MATH. India and South Africa have high values in AGR. Italy has relative high value in PHY (1.3).

The Figure 5(b) shows normalized shares of P_IM_Domestic, South Africa stands out with high value in all disciplines. Canada, Germany, France and the United Kingdom have above global average shares in most disciplines. Brazil has its highest value in PSY, the lowest value in AGR. India has below average values for most disciplines.

### 3.2.2 Effect of multi-affiliated authorship on citation impact

We fit the NBRM, to predict the effects on citations of collaborative publications, with regard to each observed factor, for G7 and BRICS countries, Table 5 mainly presents the effect of domestic multi-affiliated authorship (NM_Domestic and IM_Domestic) by disciplines.

**Table 5.** Expected percent change of citations affected by domestic multi-affiliated authorship by discipline, for (a) G7 and (b) BRICS countries, here we still use NM_mark and IM_mark, to represent NM_Domestic and IM_Domestic's participation, respectively.

**(a) G7:**

| ***Discipline*** | ***CA*** | | ***DE*** | | ***FR*** | | ***GB*** | | ***IT*** | | ***JP*** | | ***US*** | |
|---|---|---|---|---|---|---|---|---|---|---|---|---|---|---|
| | ***NM_mark*** | ***IM_mark*** | ***NM_mark*** | ***IM_mark*** | ***NM_mark*** | ***IM_mark*** | ***NM_mark*** | ***IM_mark*** | ***NM_mark*** | ***IM_mark*** | ***NM_mark*** | ***IM_mark*** | ***NM_mark*** | ***IM_mark*** |
| ***SPA*** | -14.4 | 4.1 | 18.0*** | 7.8* | -0.8 | 9.0* | -4.5 | 1.3 | 12.3* | 4.4 | 3.1 | 12.9* | 9.0** | 1.3 |
| ***NEU*** | -1.2 | -4.4 | 5.7* | 2.6 | 22.8*** | 2.3 | 15.5*** | -0.7 | 3.4 | -5.7 | 3.1 | 11.5* | 5.3*** | 1.4 |
| ***PSY*** | 18.7** | -8.8 | 3.0 | 0.3 | 12.3 | -5.4 | 20.2*** | -4.9 | 3.8 | 9.9 | -3.2 | 4.0 | 7.4** | -2.5 |
| ***IMM*** | -0.7 | -1.1 | -7.1 | 2.7 | 6.6 | 4.8 | -2.8 | -2.0 | 12.2 | -2.1 | 26.6*** | -10.0 | 7.4** | 1.8 |
| ***CLI*** | 4.9* | -11.6*** | 13.7*** | -5.8** | 11.5*** | -4.2 | 13.5*** | -1.5 | 6.8*** | -1.9 | -3.3 | 5.2 | 5.3*** | -8.2*** |
| ***PHA*** | 2.1 | -9.0 | 6.5 | 3.5 | -1.4 | 14.8* | 21.6*** | -7.6 | 10.8* | 0.9 | -5.3 | -0.6 | 6.5** | 1.2 |
| ***PHY*** | 58.7*** | 11.7*** | 19.8*** | 4.9** | 1.6 | 7.9*** | -4.6 | -2.9 | 24.3*** | 13.8*** | 36.0*** | 14.3*** | 20.0*** | 15.4*** |
| ***MOL*** | 15.8** | -7.1 | 21.0*** | 29.0*** | 5.7 | 0.6 | 20.0*** | 0.7 | 17.0** | 10.7 | 60.0*** | 3.0 | 18.8*** | 14.3*** |
| ***BIO*** | 6.1 | -6.2 | 24.3*** | 4.1 | 6.9 | -3.3 | 23.9*** | -1.1 | 3.5 | 5.2 | 11.4*** | -3.7 | 15.3*** | 0.4 |
| ***MIC*** | -0.6 | 3.3 | 10.1 | 0.5 | 10.0 | -7.6 | 13.1 | 2.1 | -3.2 | -9.7 | 9.3 | 10.5 | 5.3 | -2.4 |
| ***PLA*** | -0.8 | -7.0* | 10.4** | -1.7 | 13.0*** | -1.8 | 9.5* | 0.1 | 2.0 | -2.7 | 22.8*** | -8.1 | -0.5 | 3.2 |
| ***ENV*** | -0.2 | -1.8 | 19.3*** | 8.8** | 6.5 | 0.6 | 7.2 | 2.4 | 4.0 | -3.4 | -1.5 | -1.3 | 11.3*** | 3.6 |
| ***GEO*** | 13.5* | -0.5 | 12.1** | 6.9* | 0.9 | -0.1 | 9.0* | 11.0*** | 4.3 | -0.1 | 3.5 | 7.5 | 10.9*** | 2.9 |
| ***CHE*** | 4.6 | -4.5 | 13.4*** | 9.8*** | 6.0** | -3.0 | 7.3 | 0.1 | 8.6** | 4.0 | 52.9*** | 12.7*** | 16.8*** | 9.3*** |
| ***AGR*** | 15.6* | 18.6** | 6.6 | 6.3 | 18.4*** | 17.2** | 6.4 | -5.6 | 5.2 | 20.9** | 0.9 | -2.0 | 12.2*** | 21.3*** |
| ***MATE*** | 16.8* | -6.0 | 24.2*** | 15.7*** | 18.8*** | -0.2 | 6.8 | 3.5 | 8.7 | 8.8 | 51.0*** | 15.3*** | 24.4*** | 23.4*** |
| ***COM*** | 46.9*** | -9.3* | 5.6 | 3.5 | -13.3** | -2.5 | -19.2* | 23.7*** | 7.0 | 8.3 | -21.3* | 22.1** | 2.1 | -11.1*** |
| ***ENG*** | -0.7 | -1.2 | 50.6*** | 6.2 | 8.4** | 11.0*** | -4.2 | 12.7*** | 8.4* | -0.1 | 19.5*** | -0.3 | 11.6*** | 11.7*** |
| ***MATH*** | 16.1 | 9.8 | -20.6* | 27.3*** | -1.3 | -12.2* | 9.5 | 3.1 | -3.3 | 39.1*** | -9.5 | 19.0 | 0.2 | 17.2*** |

**(b) BRICS**

| ***Discipline*** | ***BR*** | | ***CN*** | | ***IN*** | | ***RU*** | | ***ZA*** | |
|---|---|---|---|---|---|---|---|---|---|---|
| | ***NM_mark*** | ***IM_mark*** | ***NM_mark*** | ***IM_mark*** | ***NM_mark*** | ***IM_mark*** | ***NM_mark*** | ***IM_mark*** | ***NM_mark*** | ***IM_mark*** |
| ***SPA*** | -14.0 | -5.1 | -9.1 | 25.0*** | 18.0 | 3.6 | -6.0 | 15.5* | -22.5 | 0.3 |
| ***NEU*** | 1.9 | 3.0 | 0.1 | 19.0*** | -2.8 | -4.9 | -10.7 | 69.7*** | -6.0 | 23.5 |
| ***PSY*** | 19.9 | 45.8** | 1.3 | 0.3 | -27.7 | 107.9* | 49.5 | 39.5 | -14.3 | -10.2 |
| ***IMM*** | -8.2 | 17.6 | 6.2 | 16.9** | -13.5 | 11.2 | -7.9 | -3.7 | -10.3 | 9.2 |
| ***CLI*** | 5.6* | 6.6 | 6.0*** | 11.5*** | 2.3 | -4.5 | -0.0 | 30.6** | 6.7 | -6.4 |
| ***PHA*** | -5.1 | 31.9** | 3.9 | 17.6*** | 21.8*** | 4.2 | 17.5 | 55.5** | -7.5 | -4.8 |
| ***PHY*** | -4.6 | 17.5*** | 8.9*** | 24.5*** | 4.8 | 5.2 | 16.6*** | 32.6*** | 20.4 | 18.3* |
| ***MOL*** | 17.3* | 31.2** | -5.0 | 13.8*** | 6.2 | 20.7 | 21.3* | 53.3*** | 15.8 | 89.7*** |
| ***BIO*** | 5.0 | 15.7* | 4.3* | 24.0*** | 1.0 | 3.6 | 6.9 | 18.2* | -4.6 | 14.0 |
| ***MIC*** | 8.4 | 25.6** | -2.3 | 6.6 | -8.4 | -17.3 | 5.2 | 24.2 | -11.3 | 17.1 |
| ***PLA*** | 11.7*** | 8.2 | -4.3 | 14.8*** | 12.3* | -6.6 | 12.4 | 19.5* | -0.8 | -2.5 |
| ***ENV*** | 1.6 | 10.0 | 2.6 | 9.7*** | -7.7 | -1.2 | 11.4 | 29.4* | 6.8 | -7.9 |
| ***GEO*** | -1.4 | 17.4 | -8.0*** | 9.0*** | 18.2* | 8.8 | 10.0 | 26.0*** | 11.5 | 20.7* |
| ***CHE*** | 6.5 | 19.3*** | 12.7*** | 21.4*** | 12.1*** | 20.0*** | 27.8*** | 24.4*** | 25.0* | 17.7** |
| ***AGR*** | 46.4*** | 25.4*** | -3.2 | 15.7*** | -48.7*** | 0.2 | -2.2 | 70.7** | -15.5 | 16.9 |
| ***MATE*** | 6.8 | 38.0*** | 3.4** | 21.3*** | 16.9*** | 10.4* | 43.6*** | 24.9*** | 11.0 | 13.8 |
| ***COM*** | -9.7 | 6.4 | -4.4* | 1.0 | -15.6 | -36.8*** | 14.7 | 1.5 | 62.3 | -22.0 |
| ***ENG*** | 17.9** | 5.2 | 0.0 | 5.8*** | 9.1 | -0.6 | 67.0*** | 20.4** | 12.8 | 28.0** |
| ***MATH*** | 25.9 | 7.2 | 14.7*** | 13.2*** | 24.9 | -7.8 | 11.5 | 10.9 | 31.2 | -19.3 |

*(Note: *p < 0.05; **p < 0.01; and ***p < 0.001)*

Among the significant coefficients presented in Table 5, the larger value, the higher the expected citation count for a publication. In Table 5(a), for G7 countries, regarding to the NM_Domestic authorship's coefficient, we can see there are statistically significant effect in many medicine related and biology related disciplines (e.g., for Germany, the values are 21% and 24.3% in MOL and BIO, for the United Kingdom, the values are 15.5%, 20.2%, 13.5%, 21.6%, 20% and 23.9% in the disciplines of NEU, PSY, CLI, PHA, MOL and BIO, etc.). Beyond that, we as well investigate a large coefficient in ENG (50.6%) of Germany, MATE (51%) of Japan. And the coefficient of NM_Domestic authorship is particularly large in the disciplines like PHY (58.7%) and COM (46.9%) of Canada. While we haven't investigated any significant effect for most tested units, regarding to the IM_Domestic authorship. Its positive effect mainly shows up in the disciplines like PHY, MOL, AGR and MAT among G7 countries.

In Table 5(b), for BRICS countries, far fewer citations are to be expected for a publication with the NM_Domestic authorship’s participation, than G7 countries. Especially in South Africa, we hardly investigate any obvious positive effects of the NM_Domestic authorship, on citations. But a positive effect can be still observed in countries like Brazil, China, India and Russia. Some cases stand out, e.g., with the NM_Domestic authorship, the expected citations of Brazil's publications in AGR increase by 46.4%. The expected citations of China's publications in CHE and MATH as well increase by 12.7% and 14.7%, respectively. Similar phenomena have been observed in India's PHA, Russia's CHE and ENG. Comparing with the NM_Domestic, the IM_Domestic authorship is associated with a much greater number of citations for most disciplines in Brazil, China and Russia. And we see that although South Africa doesn't show any significant association in many disciplines, with whether NM_Domestic or IM_Domestic authorship, it still has an extremely large increase (89.7%) related to the IM_Domestic authorship for publications in MOL.

By comparing NM_Domestic and IM_Domestic authorship, an interesting case is that of CLI, for Canada, Germany and the USA, the NM_Domestic authorship brings positive effects while IM_Domestic authorship brings negative effects on citations.

## 4. Conclusion and discussion

Through an exploration of collaborative publications with multi-affiliated authorship, we try to answer the questions presented previously in introduction section, mainly focus on the overview of scientific output based on collaboration, and how the two kinds of multi-affiliated authorship differently influence citation impact of collaboration in different disciplines, as well as in different sample countries.

As Hottenrott and Lawson (2017), Hottenrott, Rose and Lawson (2019), Huang and Chang (2018), Sanfilippo, Hewitt and Mackey (2018) explore in observed fields or journals, there is an increasing trend of multi-affiliation over years, we also observe a large share of collaborative publications with multi-affiliated authorship. This phenomenon as well show heterogeneities by discipline. Furthermore, we classify the multi-affiliated authorship in two types (national multi-affiliated author and international multi-affiliated author), and see medicine related and biology related disciplines have larger share of publications with the former type. Since the importance of hospital-university, a combination type of multiple affiliations, is presented by Hottenrott, Rose and Lawson (2019), with regard to medicine related research, we also take a brief look of it. From ***Figure A1*** in ***Appendix***, we can see that hospital-university/college is a frequent combination of multiple affiliations in medicine related disciplines, especially in the discipline of Clinical Medicine. While the phenomenon of Space Science's larger share for publications with international multi-affiliated authorship, may be related to the big infrastructure collaboration all over the world.

Previous studies (Hottenrott & Lawson, 2017; Huang & Chang, 2018; Sanfilippo, Hewitt, & Mackey, 2018) indicate that multi-affiliated authorship play a positive role on citation impact. We here employ NBRM, no merely observe the same phenomenon, but also see how different national and international multi-affiliated author effect on citation impact across fields. We find that in medicine related and biology related disciplines, the national multi-affiliated authors are associated with more citations than the international multi-affiliated author, while Space Science, Geosciences and Mathematics show up the opposite phenomenon, the international multi-affiliated authors relate to more citations.

We go further, try to explore their effect on citation impact for different S&T level countries. It is worth mentioning that we also distinguish domestic multi-affiliated authorship and foreign multi-affiliated authorship regarding to each specific country, taking their different role into account, here we focus on the domestic part. We find

France has a very high share of publications with domestic national multi-affiliated authorship, with China and Russia following behind. For publications with domestic international multi-affiliated authorship, South Africa has the largest share among G7 and BRICS countries. Relating to this phenomenon, science policy might be a driving force. We find some connotations displayed in the human resources related documents of the French National Centre for Scientific Research (CNRS)[1], the institution contributes mostly multi-affiliated publications (in our dataset) for France. Similar implications can be found in policies of foundation institutions of South Africa, e.g. the National Research Foundation (NRF)[2].

We as well observe an interesting result in citation effect section, for most disciplines of G7 countries, the domestic national multi-affiliated authorship relates to more on citation impact, while domestic international multi-affiliated authorship is more positively influential in most BRICS countries. We investigate the affiliation links for BRICS countries, and find that more foreign affiliations come from G7 countries. Citation impact can be increased by collaborating with high R&D intensity or high S&T level countries (Bordons, Aparicio, & Costas, 2013; Bordons, Gonzalez-Albo, Aparicio, & Moreno, 2015; Bote, Olmeda-Gomez, & de Moya-Anegon, 2013; Glänzel, 2001), in our study, it has been explored that for BRICS countries, constructing affiliation links to high S&T level countries may also bring benefit to the impact of scientific output.

[1] http://www.cnrs.fr/en/science-news/docs/HRS4R-en.pdf: HUMAN RESOURCES STRATEGY FOR RESEARCHERS.

[2] https://www.nrf.ac.za/information-resources/annual-performance-plans: Annual Performance Plan.

Nevertheless, this study currently has several limitations. Discipline schema is one of them, we here use ESI category, the granularity may be too thick to observe some special fields which are not listed in this category. Another is that we only consider a short citation period (3-year citation window) in this study, the effect of multi-affiliated authorships on long-term citations needs further investigation.

Despite these limitations above, our study enables an analysis of multi-affiliated researchers' effect on the scientific output of research collaboration, from 19 ESI disciplines, demonstrating their positive influence presented by ESF (2013), in facilitating cooperation. Multiple affiliation happening among one country or multiple countries are different, we therefore classify multiple affiliation by their affiliation combination, from national face or international face, investigate and compare these two multiple affiliation types in scientific production of research collaboration and effect on citation impact, fill the blank of research studying how citations are influenced by national versus international multi-affiliated authorship across science fields. Like general collaboration patterns linking to countries' different conditions (Garg, Kumar, & Bebi, 2018; Maisonobe, Eckert, Grossetti, Jegou, & Milard, 2016), we also take country's developing level into account, to see the share of production with the two types of multi-affiliated authorship, as well as which types are more influential for citation impact in G7 and BRICS countries across science fields. Considering different multiple affiliations links might have different influences on citation impact, we attempt to explore how scientific combinations happen among countries or institutions by multi-affiliated researchers, in our further research.

4(1), 1-13.

## 6. Appendix

**Table A1.** Share of publications with multi-affiliated authorship by ESI disciplines.

| *Field* | *Discipline (Abbreviation)* | *P_NM* | *P_IM* |
|---|---|---|---|
| ***Space Science*** | ***SPA*** | 41.9% | 33.9% |
| ***Medicine related*** | ***NEU*** | 51.3% | 16.8% |
| | ***PSY*** | 50.3% | 14.5% |
| | ***IMM*** | 48.3% | 15.9% |
| | ***CLI*** | 45.0% | 10.3% |
| | ***PHA*** | 37.1% | 11.4% |
| ***Physics*** | ***PHY*** | 35.0% | 20.4% |
| ***Biology related*** | ***MOL*** | 45.9% | 17.5% |
| | ***BIO*** | 39.7% | 14.7% |
| | ***MIC*** | 39.2% | 14.9% |
| | ***PLA*** | 29.6% | 14.8% |
| ***Environment/Ecology*** | ***ENV*** | 31.5% | 17.8% |
| ***Geosciences*** | ***GEO*** | 30.2% | 18.0% |
| ***Chemistry*** | ***CHE*** | 32.2% | 13.9% |
| ***Agricultural Sciences*** | ***AGR*** | 27.1% | 12.2% |
| ***Engineering related*** | ***MATE*** | 29.8% | 14.3% |
| | ***COM*** | 21.6% | 12.4% |
| | ***ENG*** | 20.5% | 11.7% |
| ***Mathematics*** | ***MATH*** | 15.8% | 10.7% |

**Table A2.** Number of publications containing NM_Domestic, by discipline, for each country's Top 3 (ranked with regard to Publications with NM_Domestic) institutions.

| | Country | Institution | Publications with NM_Domestic | Share (in total) |
|---|---|---|---|---|
| G7 | Canada | UNIVERSITY OF TORONTO | 1472 | 8.0% |
| | | UNIVERSITY OF BRITISH COLUMBIA | 1073 | 9.7% |
| | | MCGILL UNIVERSITY | 732 | 7.6% |
| | Germany | MAX PLANCK SOCIETY | 3073 | 14.4% |
| | | HELMHOLTZ ASSOCIATION | 2331 | 21.9% |
| | | UNIVERSITY OF MUNICH | 1267 | 14.9% |
| | France | CENTRE NATIONAL DE LA RECHERCHE SCIENTIFIQUE (CNRS) | 8951 | 42.8% |
| | | INSTITUT NATIONAL DE LA SANTE ET DE LA RECHERCHE MEDICALE (INSERM) | 3629 | 39.0% |
| | | PIERRE & MARIE CURIE UNIVERSITY - PARIS 6 | 3412 | 27.2% |
| | United Kingdom | UNIVERSITY COLLEGE LONDON | 1760 | 10.8% |
| | | IMPERIAL COLLEGE LONDON | 1297 | 10.2% |
| | | UNIVERSITY OF OXFORD | 1289 | 8.4% |
| | Italy | ISTITUTO NAZIONALE DI FISICA NUCLEARE | 2835 | 53.9% |
| | | CONSIGLIO NAZIONALE DELLE RICERCHE (CNR) | 1862 | 17.0% |
| | | UNIVERSITY OF MILAN | 1271 | 15.8% |
| | Japan | UNIVERSITY OF TOKYO | 1283 | 7.9% |
| | | JAPAN SCIENCE & TECHNOLOGY AGENCY (JST) | 1268 | 19.1% |
| | | RIKEN | 701 | 13.8% |
| | USA | HARVARD UNIVERSITY | 4504 | 12.5% |
| | | BRIGHAM AND WOMEN'S HOSPITAL | 1678 | 26.1% |
| | | STANFORD UNIVERSITY | 1176 | 8.2% |
| BRICS | Brazil | UNIVERSIDADE DE SAO PAULO | 829 | 4.9% |
| | | UNIVERSIDADE FEDERAL DO RIO DE JANEIRO | 436 | 8.4% |
| | | UNIVERSIDADE FEDERAL DO RIO GRANDE DO SUL | 356 | 7.1% |
| | China | CHINESE ACADEMY OF SCIENCES | 4900 | 7.5% |
| | | PEKING UNIVERSITY | 957 | 7.1% |
| | | SHANGHAI JIAO TONG UNIVERSITY | 825 | 6.1% |
| | India | COUNCIL OF SCIENTIFIC & INDUSTRIAL RESEARCH (CSIR) - INDIA | 162 | 4.2% |
| | | TATA INSTITUTE OF FUNDAMENTAL RESEARCH | 94 | 7.1% |
| | | INDIAN INSTITUTE OF SCIENCE (IISC) - BANGLORE | 92 | 3.5% |
| | Russia | RUSSIAN ACADEMY OF SCIENCES | 3636 | 13.7% |
| | | NOVOSIBIRSK STATE UNIVERSITY | 886 | 30.9% |
| | | LOMONOSOV MOSCOW STATE UNIVERSITY | 867 | 13.1% |
| | South Africa | UNIVERSITY OF CAPE TOWN | 442 | 12.7% |
| | | STELLENBOSCH UNIVERSITY | 350 | 13.6% |
| | | UNIVERSITY OF WITWATERSRAND | 328 | 14.2% |

*(Note: Share (in total) is share of Publications with NM_Domestic, account for total institutionally collaborative publications.)*

**Table A3.** Number of publications containing IM_Domestic, by discipline, for each country's Top 3 (ranked with regard to Publications with IM_Domestic) institutions.

| | Country | Institution | Publications with IM_Domestic | Share (in total) |
|---|---|---|---|---|
| G7 | Canada | UNIVERSITY OF TORONTO | 3102 | 16.9% |
| | | UNIVERSITY OF BRITISH COLUMBIA | 1868 | 16.9% |
| | | MCGILL UNIVERSITY | 1738 | 18.1% |
| | Germany | MAX PLANCK SOCIETY | 6962 | 32.6% |
| | | HELMHOLTZ ASSOCIATION | 1686 | 15.8% |
| | | RUPRECHT KARL UNIVERSITY HEIDELBERG | 1482 | 16.8% |
| | France | CENTRE NATIONAL DE LA RECHERCHE SCIENTIFIQUE (CNRS) | 3157 | 15.1% |
| | | PIERRE & MARIE CURIE UNIVERSITY - PARIS 6 | 2106 | 16.8% |
| | | UNIVERSITY OF PARIS SUD - PARIS XI | 1107 | 13.3% |
| | United Kingdom | UNIVERSITY OF OXFORD | 3839 | 25.1% |
| | | UNIVERSITY COLLEGE LONDON | 3560 | 21.8% |
| | | UNIVERSITY OF CAMBRIDGE | 3534 | 23.9% |
| | Italy | ISTITUTO NAZIONALE DI FISICA NUCLEARE | 1013 | 19.3% |
| | | SAPIENZA UNIVERSITY ROME | 984 | 10.3% |
| | | UNIVERSITY OF PADUA | 858 | 11.1% |
| | Japan | UNIVERSITY OF TOKYO | 1781 | 11.0% |
| | | TOHOKU UNIVERSITY | 1162 | 12.7% |
| | | KYOTO UNIVERSITY | 926 | 8.0% |
| | United States | HARVARD UNIVERSITY | 6824 | 18.9% |
| | | UNIVERSITY OF CALIFORNIA BERKELEY | 2240 | 17.4% |
| | | STANFORD UNIVERSITY | 2192 | 15.2% |
| BRICS | Brazil | UNIVERSIDADE DE SAO PAULO | 1870 | 11.1% |
| | | UNIVERSIDADE ESTADUAL DE CAMPINAS | 476 | 8.7% |
| | | UNIVERSIDADE ESTADUAL PAULISTA | 435 | 6.9% |
| | China | CHINESE ACADEMY OF SCIENCES | 7533 | 11.6% |
| | | ZHEJIANG UNIVERSITY | 1997 | 14.8% |
| | | SHANGHAI JIAO TONG UNIVERSITY | 1917 | 14.3% |
| | India | INDIAN INSTITUTE OF SCIENCE (IISC) - BANGLORE | 337 | 12.8% |
| | | COUNCIL OF SCIENTIFIC & INDUSTRIAL RESEARCH (CSIR) - INDIA | 227 | 5.8% |
| | | INDIAN INSTITUTE OF TECHNOLOGY (IIT) - BOMBAY | 168 | 10.9% |
| | Russia | RUSSIAN ACADEMY OF SCIENCES | 3430 | 13.0% |
| | | LOMONOSOV MOSCOW STATE UNIVERSITY | 1053 | 15.9% |
| | | SAINT PETERSBURG STATE UNIVERSITY | 549 | 19.0% |
| | South Africa | UNIVERSITY OF KWAZULU NATAL | 1010 | 40.6% |
| | | UNIVERSITY OF CAPE TOWN | 988 | 28.5% |
| | | UNIVERSITY OF PRETORIA | 712 | 30.4% |

*(Note: Share (in total) is share of Publications with IM_Domestic, account for total institutionally collaborative publications.)*

**Table A4.** P_NM share by discipline, for G7 and BRICS countries

| NM | G7 | | | | | | | BRICS | | | | |
|---|---|---|---|---|---|---|---|---|---|---|---|---|
| | CA | DE | FR | GB | IT | JP | US | BR | CN | IN | RU | ZA |
| Space Science | 12.2% | 14.1% | 29.2% | 8.3% | 33.0% | 23.4% | 23.5% | 13.5% | 26.5% | 15.4% | 26.7% | 20.2% |
| Neuroscience & Behavior | 30.6% | 36.5% | 58.5% | 21.5% | 43.3% | 37.8% | 36.6% | 37.0% | 32.0% | 26.9% | 30.9% | 15.6% |
| Psychiatry/Psychology | 36.4% | 30.8% | 57.5% | 23.6% | 35.1% | 35.6% | 35.9% | 39.3% | 31.3% | 18.5% | 20.0% | 19.5% |
| Immunology | 33.1% | 28.4% | 56.7% | 21.2% | 30.3% | 29.8% | 31.7% | 39.9% | 34.5% | 21.2% | 26.9% | 31.2% |
| Clinical Medicine | 29.6% | 26.9% | 54.2% | 25.3% | 31.3% | 27.1% | 34.4% | 35.3% | 31.1% | 26.5% | 28.2% | 27.2% |
| Pharmacology & Toxicology | 26.2% | 20.6% | 48.9% | 16.4% | 26.8% | 27.5% | 25.3% | 29.9% | 36.5% | 20.8% | 30.4% | 17.4% |
| Physics | 16.6% | 21.5% | 27.4% | 8.6% | 43.8% | 25.6% | 16.2% | 16.9% | 35.0% | 14.6% | 37.9% | 18.9% |
| Molecular Biology & Genetics | 22.4% | 28.3% | 53.9% | 17.6% | 36.1% | 32.1% | 28.8% | 29.9% | 33.3% | 22.4% | 46.8% | 17.4% |
| Biology & Biochemistry | 20.0% | 25.8% | 49.7% | 14.7% | 30.7% | 32.2% | 25.0% | 28.9% | 35.3% | 21.5% | 39.1% | 13.4% |
| Microbiology | 24.3% | 22.5% | 52.0% | 14.1% | 21.7% | 25.3% | 23.3% | 28.2% | 40.8% | 17.4% | 26.6% | 25.2% |
| Plant & Animal Science | 15.6% | 17.3% | 31.7% | 12.6% | 13.1% | 23.0% | 19.2% | 18.8% | 36.1% | 18.8% | 20.7% | 26.4% |
| Environment/Ecology | 15.2% | 22.3% | 33.1% | 12.1% | 13.9% | 14.8% | 19.6% | 21.8% | 33.8% | 14.5% | 19.5% | 21.9% |
| Geosciences | 9.9% | 14.5% | 25.8% | 11.0% | 19.1% | 17.4% | 18.2% | 15.4% | 31.3% | 11.1% | 27.5% | 15.0% |
| Chemistry | 10.3% | 19.5% | 33.2% | 8.0% | 26.1% | 29.9% | 14.5% | 19.6% | 38.4% | 17.2% | 36.3% | 11.7% |
| Agricultural Sciences | 14.4% | 14.7% | 38.9% | 10.4% | 13.5% | 20.5% | 12.3% | 12.0% | 36.9% | 35.2% | 17.8% | 26.9% |
| Materials Science | 10.1% | 21.7% | 27.4% | 8.0% | 25.2% | 20.9% | 14.7% | 18.6% | 34.1% | 15.0% | 27.9% | 25.8% |
| Computer Science | 5.5% | 10.4% | 26.4% | 3.8% | 12.5% | 11.0% | 8.2% | 14.7% | 27.2% | 6.9% | 27.7% | 23.8% |
| Engineering | 7.4% | 11.6% | 24.2% | 5.2% | 11.3% | 9.5% | 8.9% | 14.5% | 26.8% | 8.9% | 29.3% | 8.7% |
| Mathematics | 2.4% | 4.3% | 19.7% | 2.4% | 5.3% | 5.2% | 3.9% | 3.4% | 22.1% | 4.4% | 31.6% | 2.7% |

**Table A5.** P_IM share by discipline, for G7 and BRICS countries

| IM | G7 | | | | | | | BRICS | | | | |
|---|---|---|---|---|---|---|---|---|---|---|---|---|
| | CA | DE | FR | GB | IT | JP | US | BR | CN | IN | RU | ZA |
| Space Science | 28.9% | 38.8% | 33.5% | 35.7% | 31.8% | 24.7% | 27.5% | 35.6% | 32.8% | 15.6% | 40.0% | 50.0% |
| Neuroscience & Behavior | 21.7% | 26.4% | 21.3% | 33.9% | 20.6% | 14.1% | 18.8% | 20.7% | 23.3% | 11.7% | 41.1% | 45.9% |
| Psychiatry/Psychology | 18.0% | 24.8% | 19.6% | 26.9% | 23.8% | 16.5% | 13.4% | 22.9% | 19.4% | 16.0% | 44.6% | 33.5% |
| Immunology | 17.5% | 18.7% | 18.4% | 27.8% | 12.9% | 12.4% | 18.5% | 14.7% | 20.0% | 15.4% | 20.5% | 42.0% |
| Clinical Medicine | 16.4% | 16.2% | 12.4% | 19.0% | 12.2% | 8.3% | 13.1% | 11.8% | 15.0% | 9.5% | 13.4% | 29.3% |
| Pharmacology & Toxicology | 18.3% | 17.0% | 15.6% | 18.9% | 10.5% | 9.6% | 15.9% | 10.1% | 12.4% | 9.3% | 15.1% | 29.4% |
| Physics | 31.2% | 32.4% | 25.6% | 32.2% | 24.9% | 16.2% | 26.8% | 26.3% | 19.3% | 12.1% | 29.0% | 41.2% |
| Molecular Biology & Genetics | 18.3% | 21.8% | 21.4% | 28.3% | 19.9% | 15.3% | 21.8% | 19.7% | 23.3% | 13.9% | 23.8% | 39.6% |
| Biology & Biochemistry | 20.6% | 22.8% | 19.9% | 26.1% | 13.5% | 11.9% | 20.1% | 14.2% | 17.6% | 12.5% | 23.1% | 41.9% |
| Microbiology | 20.6% | 18.9% | 18.1% | 25.2% | 14.0% | 17.5% | 18.2% | 15.2% | 16.0% | 10.6% | 19.7% | 38.4% |
| Plant & Animal Science | 19.8% | 22.8% | 26.0% | 26.8% | 13.3% | 12.0% | 15.8% | 9.1% | 16.7% | 11.4% | 18.0% | 34.9% |
| Environment/Ecology | 22.9% | 26.4% | 29.5% | 30.8% | 16.9% | 16.5% | 18.4% | 16.9% | 19.0% | 15.1% | 16.3% | 33.2% |
| Geosciences | 20.1% | 24.9% | 26.5% | 26.2% | 14.7% | 15.2% | 18.6% | 12.6% | 21.2% | 10.7% | 15.6% | 36.2% |
| Chemistry | 25.1% | 22.5% | 21.4% | 23.7% | 15.6% | 13.6% | 21.4% | 14.1% | 13.4% | 9.6% | 13.2% | 32.9% |
| Agricultural Sciences | 21.2% | 19.0% | 21.1% | 23.0% | 9.5% | 12.9% | 16.2% | 4.6% | 18.7% | 8.3% | 16.1% | 29.0% |
| Materials Science | 22.6% | 22.8% | 21.2% | 27.2% | 16.0% | 17.2% | 24.9% | 13.8% | 15.9% | 11.2% | 16.3% | 35.2% |
| Computer Science | 19.0% | 16.1% | 17.5% | 20.6% | 12.6% | 13.4% | 14.1% | 10.9% | 14.2% | 5.8% | 26.3% | 24.5% |
| Engineering | 19.3% | 15.4% | 16.5% | 21.3% | 12.9% | 11.1% | 14.6% | 7.9% | 15.9% | 5.8% | 16.9% | 27.2% |
| Mathematics | 15.3% | 12.3% | 11.8% | 14.3% | 8.0% | 6.6% | 12.1% | 8.4% | 11.6% | 5.7% | 21.7% | 35.3% |

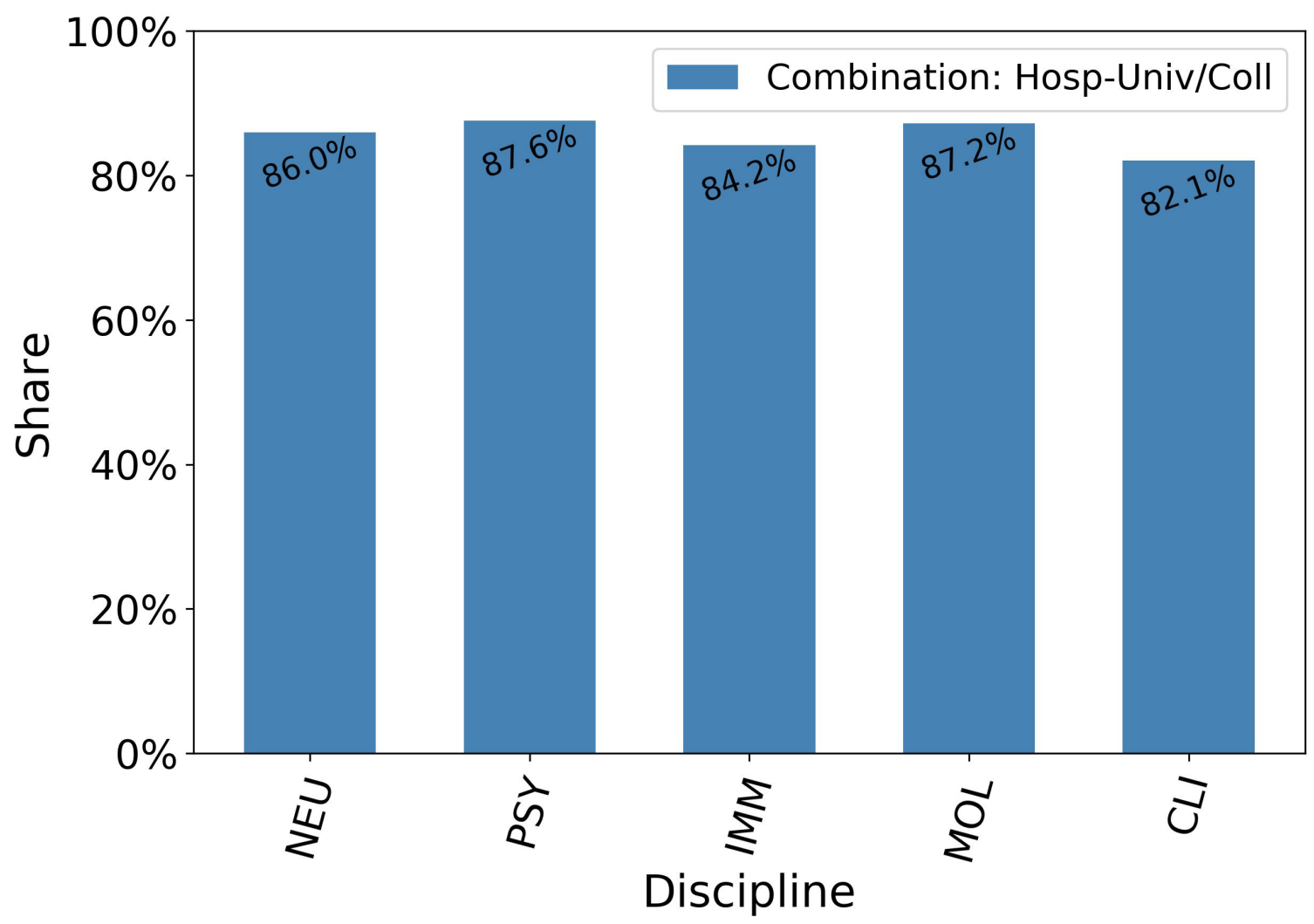

**Figure A1.** The combination of *Hosp-Univ/Coll* means the multiple affiliations combination between hospital type affiliations and university or college type affiliations. The bars present share of publications with the Hosp-Univ/Coll combination, in publications with multi-affiliated authors from hospital, of 5 frequent P_NM disciplines.